\documentclass[aps,showpacs,twocolumn,prb]{revtex4-1}

\usepackage[dvipdfmx]{graphicx}
\usepackage{dcolumn}
\usepackage{bm}
\usepackage{hyperref}
\usepackage{breakurl}
\usepackage[utf8]{inputenc}
\usepackage{pslatex}
\usepackage{textcomp} 
\usepackage{color}
\usepackage{lmodern} 
\usepackage{amsmath}
\usepackage{amsfonts}
\usepackage{amssymb}

\graphicspath{{./figures/}}

\hypersetup{
    breaklinks = true,
    colorlinks=true,
    linkcolor = blue,
    citecolor = blue,    
    filecolor=black,
	urlcolor = magenta
}

\begin{document}

\title{Magnetocrystalline anisotropy of Fe$_5$PB$_2$ and its alloys with Co and 5$\boldsymbol{d}$ elements:\\ a combined first-principles and experimental study}

\author{Miros\l{}aw Werwi\'nski}\email[Corresponding author: ]{werwinski@ifmpan.poznan.pl}
\affiliation{Institute of Molecular Physics, Polish Academy of Sciences, M. Smoluchowskiego 17, 60-179 Pozna\'{n}, Poland}

\author{Alexander Edstr\"{o}m}
\affiliation{Materials Theory, ETH Z\"{u}rich, Wolfgang-Pauli-Str. 27, 8093 Z\"{u}rich, Switzerland}

\author{J\'{a}n Rusz}
\affiliation{Department of Physics and Astronomy, Uppsala University, Box 516, SE-751 20 Uppsala, Sweden}

\author{Daniel Hedlund}
\author{Klas Gunnarsson}
\author{Peter Svedlindh}
\affiliation{Department of Engineering Sciences, Uppsala University, Box 534, SE-751 21 Uppsala, Sweden}

\author{Johan Cedervall}
\author{Martin Sahlberg}
\affiliation{Department of Chemistry -- The \AA{}ngstr\"{o}m Laboratory, Uppsala University, Box 538, SE-751 21 Uppsala, Sweden}

\date{\today}

\newcommand{\fefivep}{Fe$_5$PB$_2$}
\newcommand{\fefivesi}{Fe$_5$SiB$_2$}

\newcommand{\cofivep}{Co$_5$PB$_2$}
\newcommand{\cofivesi}{Co$_5$SiB$_2$}

\newcommand{\fecofivemax}{(Fe$_{0.85}$Co$_{0.15}$)$_5$PB$_2$}

\newcommand{\fecofivep}{(Fe$_{1-x}$Co$_{x}$)$_5$PB$_2$}
\newcommand{\fecofivesi}{(Fe$_{1-x}$Co$_{x}$)$_5$SiB$_2$}
\newcommand{\fecotwo}{(Fe$_{1-x}$Co$_{x}$)$_2$B}

\begin{abstract}

The Fe$_5$PB$_2$ compound  offers tunable magnetic properties via the possibility of various combinations of substitutions on the Fe and P-sites. 
Here, we present a combined computational and experimental study of the magnetic properties of \fecofivep{}.
Computationally, we are able to explore the full concentration range, 
while the real samples were only obtained for $0 \leq x \leq 0.7$.
The calculated magnetic moments, Curie temperatures, and magnetocrystalline anisotropy energies (MAEs) are found to decrease with increasing  Co concentration. 
Co substitution allows for tuning the Curie temperature in a wide range of values, from about six hundred to zero kelvins. 
As the MAE depends on the electronic structure in the vicinity of Fermi energy, the geometry of the Fermi surface of \fefivep{} and the \textbf{k}-resolved contributions to the MAE are discussed. 
Low temperature measurements of an effective anisotropy constant for a series of \fecofivep{} samples determined the highest value of 0.94~MJ\,m$^{-3}$ for the terminal \fefivep{} composition, which then decreases with increasing Co concentration, thus confirming the computational result that Co alloying of \fefivep{} is not a good strategy to increase the MAE of the system.
However, the relativistic version of the fixed spin moment method reveals that a reduction in the magnetic moment of \fefivep{}, by about 25\%, produces a fourfold increase of the MAE.
Furthermore, calculations for (Fe$_{0.95}$X$_{0.05}$)$_5$PB$_2$ (X = 5$d$ element) indicate that 5\% doping of \fefivep{} with W or Re should double the MAE. 
These are results of high interest for, e.g., permanent magnet applications, where a large MAE is crucial.

\end{abstract}


\maketitle

\section{Introduction}
\begin{sloppypar} 
%
%

Many sectors of modern technology depend on magnetic materials,
which are used in such ubiquitous applications as electric motors, power generators, transformers, and recording media.
Hence, magnetic materials are crucial, not only for the digital technology revolution observed in past decades, but also for the green energy revolution expected within the years to come.
The fundamentally and technologically most important intrinsic parameters of magnetic materials include the Curie temperature ($T_\mathrm{C}$), saturation magnetization ($M_\mathrm{s}$), and magnetocrystalline anisotropy energy (MAE).
These parameters are important in a wide variety of applications, including hard and soft magnetic materials for energy conversion, spintronics, and information storage.
Thus, the ability to predict these basic magnetic parameters from first principles is of utmost importance, and accurate modern electronic structure calculations provide an indispensable tool for exploring new materials with desired properties. 
In parallel, experimental synthesis and characterization retains its fundamental importance and a close interplay between computational and experimental work is of ever increasing value in modern materials discovery.

One example of an area in which the search for new magnetic materials, with specific combinations of properties, has been intense in recent years is that of permanent magnets. In this field it is typically desirable to have large $M_\mathrm{s}$, $T_\mathrm{C}$ and MAE. This combination is obtained in the commonly used rare-earth transition metal compounds, such as NdFe$_{14}$B$_2$. However, the so called \textit{Rare-Earth Crisis}~\cite{bourzac_rare-earth_2011} 
triggered immense international research initiatives in search for new substitute permanent magnet materials with reduced amounts of, or no, rare-earth elements~\cite{niarchos_toward_2015, hirosawa_current_2015, skomski_magnetic_2016, li_recent_2016,hirosawa_perspectives_2017}. The main challenge in this context is obtaining a sufficiently large MAE in transition metal compounds, where a uniaxial (e.g. tetragonal or hexagonal) crystal structure is a crucial prerequisite. Other areas of applications depend upon other combinations of properties. For example, for magnetocaloric solid state cooling, it is desirable to be able to tune the ordering temperature such that it coincides with the operating temperature (often room temperature)\cite{gutfleisch_magnetic_2011,fahler_caloric_nodate}. 

Various works have shown how strain engineering or alloying can be used to carefully tune the properties of magnetic materials to obtain desired functionality. For example, it was shown that a careful control of strain and alloy concentration allows for a large MAE in bct FeCo alloys~\cite{burkert_giant_2004,andersson_perpendicular_2006,kota_degree_2012,turek_magnetic_2012}. The potential route to FeCo-based permanent magnets offered by that work inspired subsequent studies aiming to stabilize tetragonality in FeCo by B or C-impurities~\cite{delczeg-czirjak_stabilization_2014,reichel_increased_2014,reichel_soft_2015}. Also the tetragonal (Fe$_{1-x}$Co$_x$)$_2$B compound has been carefully studied due to its tunable MAE as function of $x$~\cite{kuzmin_towards_2014,belashchenko_origin_2015,dane_density_2015,edstrom_magnetic_2015,wallisch_synthesis_2015}  which, furthermore, has an intriguing temperature dependence~\cite{iga_magnetocrystalline_1970,zhuravlev_spin-fluctuation_2015,edstrom_magnetic_2015}. It was 
also shown, in both calculations and experiments, that small amounts of 5$d$ substitutions on the Fe/Co site allowed a large increase in the MAE of this material~\cite{edstrom_magnetic_2015}.  

The tetragonal family of compounds with compositions (Fe$_{1-x}$Co$_{x}$)$_5$P$_{1-y}$Si$_y$B$_2$ has also been the subject of numerous recent studies~\cite{mcguire_magnetic_2015,werwinski_magnetic_2016,cedervall_magnetostructural_2016,lamichhane_study_2016,hedlund_magnetic_2017,cedervall_influence_2018}. Additionally, other chemical substitutions, including Mn on the Fe/Co site~\cite{mcguire_magnetic_2015}, have been considered. Due to the broad range of chemical compositions available, this material offers wide tunability of its magnetic properties. Furthermore, the tetragonal crystal structure could potentially allow for a large MAE and, thus, make the compounds interesting within the context of permanent magnet applications. The materials also exhibit other interesting aspects, such as the temperature dependent spin-reorientation transition in Fe$_5$SiB$_2$~\cite{cedervall_magnetostructural_2016}.

%
The aim of the work is to investigate the effect of the Co and 5$d$ dopants on the tunable magnetic properties of the technologically promising semi-hard  Fe$_5$PB$_2$ compound.
%
%
\begin{figure}[ht]
\includegraphics[width=0.65\columnwidth]{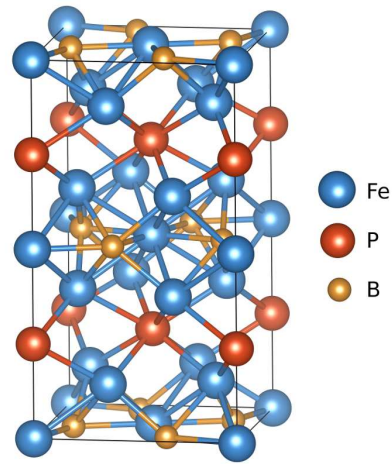}
\caption{\label{fig:structure}
The crystal structure of \fefivep{}, space group $I$4/$mcm$ (no. 140).
}
\end{figure}
\fefivep{} crystallizes in the Cr$_5$B$_3$-type structure with a body-centered tetragonal (bct) unit cell, space group $I$4/$mcm$~\cite{rundqvist_x-ray_1962} (see Fig.~\ref{fig:structure}).
The unit cell of \fefivep{} consists of 4 formula units (32 atoms).
Fe atoms occupy two inequivalent sites Fe$_1$ (16$l$) and Fe$_2$ (4$c$).
Fe$_1$ atoms are distributed on the 16-fold position, Fe$_2$ and P on the 4-fold, and B on the 8-fold position.

%
One of the motivations to investigate the \fecofivep{} system are our previous results for isostructural \fecofivesi{} system (with Si in place of P), 
for which we have predicted the highest MAE~=~1.16~MJ\,m$^{-3}$ for Co concentration $x$~=~0.3.~\cite{werwinski_magnetic_2016}
Next, the (Fe$_{0.8}$Co$_{0.2}$)$_5$SiB$_2$ sample (with Co concentration $x$~=~0.2) was synthesized by McGuire and Parker~\cite{mcguire_magnetic_2015} and
their magnetic measurements showed an increase of the anisotropy field after Co substitution, which supports our prediction.
All the previous experimental studies conducted on the \fecofivesi{} system are limited to the Fe-rich compositions, 
while Co$_5$SiB$_2$ is not known to form.~\cite{mcguire_magnetic_2015}
For melt-spun samples Fe$_5$(Si$_{0.75}$Ge$_{0.25}$)B$_2$
Lejeune et al. were determined a relatively high anisotropy constant $K_\mathrm{1}$ of about 0.5~MJ\,m$^{-3}$ at room temperature, which is about double the value for \fefivesi{}.~\cite{lejeune_synthesis_2018,werwinski_magnetic_2016}
Recently, we also presented a combined experimental and theoretical study of the Fe$_5$Si$_{1-x}$P$_{x}$B$_2$ system, which showed the highest  anisotropy constant for the terminal \fefivep{} composition.~\cite{hedlund_magnetic_2017}

%
\fefivep{} has high $T_\mathrm{C}$ of about 655\textpm2~K, magnetic moment of 1.72~$\mu_{\mathrm{B}}$/Fe atom (8.60~$\mu_{\mathrm{B}}$/f.u.), and anisotropy constant $K_\mathrm{1}$ of 0.50~MJ\,m$^{-3}$ measured at 2~K for single crystal.~\cite{lamichhane_study_2016}
The value of an effective anisotropy constant $K_{\textrm{eff}}$ of \fefivep{} obtained in our previous work is however significantly higher and equal to $\sim$0.9~MJ\,m$^{-3}$ at 10~K.~\cite{hedlund_magnetic_2017} 
An important parameter, in context of permanent magnets, is magnetic hardness, defined as:
\begin{equation}
 \kappa = \sqrt{\frac{|K|}{\mu_0 M_\mathrm{S}^2}},
\end{equation}
where $K$ is the magnetic anisotropy constant and $M_\mathrm{S}$ is the saturation magnetization.
An empirical rule $\kappa > 1$ specifies whether the material have a chance to resist self-demagnetization.~\cite{skomski_magnetic_2016}
From the experimental values of $K_{\textrm{eff}} \sim 0.65$~MJ\,m$^{-3}$ and $M_\mathrm{S}=0.87$~MA/m~\cite{hedlund_magnetic_2017}, we determined for \fefivep{} $\kappa = 0.69$ (at 300~K).
It implies, that without a further engineering of the anisotropy constant,
 \fefivep{} will stay in a category of semi-hard magnets.~\cite{skomski_magnetic_2016}

In this work we consider alloying of \fefivep{} with Co and 5$d$ elements. 
In our recent study of \fecofivep{} alloys we observed a reduction in magnetization and Curie temperature with an increase of Co concentration.~\cite{cedervall_influence_2018}
McGuire and Parker also found that 20\% Co alloying in \fecofivep{} leads to decrease in magnetization, Curie temperature and anisotropy field.~\cite{mcguire_magnetic_2015}
%
%
Previously we showed also that increase of the MAE of 3$d$ alloys can be achieved through doping with 5$d$ elements.~\cite{edstrom_magnetic_2015}
In this work we follow this idea and calculate the resultant MAEs of \fefivep{}-based alloys with 5\% substitutions of each 5$d$ element in place of Fe.

\section{Computational and Experimental Details}\label{sec:exp_comp_details}
\subsection{Computational Details}\label{subsec:comp_details}
%
%
The electronic band structure calculations for \fecofivep{} and (Fe$_{0.95}$X$_{0.05}$)$_5$PB$_2$ (X = 5$d$ element) systems
were carried out with use of the 
full-potential local-orbital electronic structure code FPLO14.0-49~\cite{koepernik_full-potential_1999} using a fixed atomic-like basis set.
The FPLO was an optimal choice for the accurate calculations of MAE
due to
the full potential and fully relativistic character of the code.
To model the Co and 5$d$ alloying we used the supercell method.
The generalized gradient approximation (GGA) was used in the Perdew-Burke-Ernzerhof form (PBE).~\cite{perdew_generalized_1996}
A $16 \times 16 \times 16$ \textbf{k}-mesh was found to lead to well converged results of the MAE.
For \textbf{k}-point integration, the tetrahedron method was used.~\cite{blochl_improved_1994}
The energy and charge density convergence criteria of $\sim 10^{-7}$~eV and 10$^{-6}$, respectively, were applied simultaneously.
%
%
The lattice parameters and Wyckoff positions were optimized for \fefivep{} and \cofivep{} within a spin-polarized scalar-relativistic approach.
The crystallographic parameters for compositions with intermediate Co concentrations were taken from calculations of full lattice relaxation carried out previously in virtual crystal approximation.~\cite{cedervall_influence_2018}
For the (Fe$_{0.95}$X$_{0.05}$)$_5$PB$_2$ supercells we used the same crystallographic parameters as for the \fefivep{}.
%
%
The MAE was evaluated as a difference between the fully relativistic total energies calculated for quantization axes [100] and [001].
In the adopted sign convention the positive sign of MAE corresponds to an easy magnetization axis along the [001] direction.
%
%
The Fermi surface (FS) of \fefivep{} was calculated on a 28$^3$ $\mathbf{k}$-mesh in a boundary of the first Brillouin zone.
%
%
Using the fully relativistic fixed spin moment (FSM) scheme~\cite{schwarz_itinerant_1984} we study the MAE as a function of total magnetic moment ($m$) for \fefivep{}.
%
%
\begin{figure}[ht]
\includegraphics[width=0.95\columnwidth]{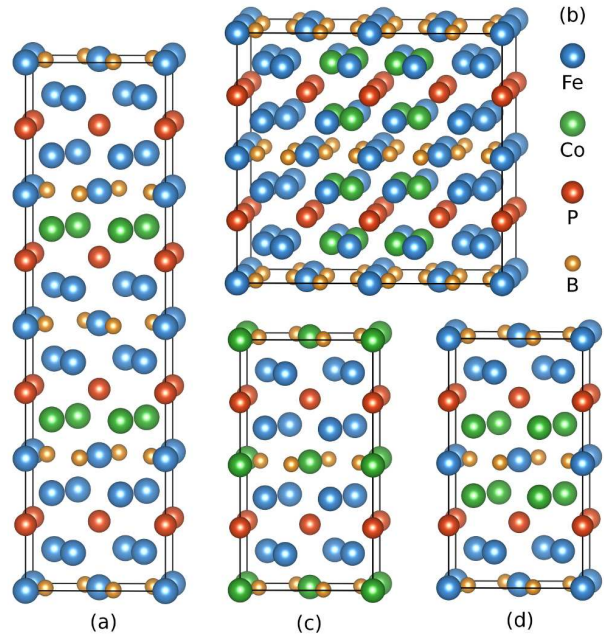}
\caption{\label{fig:supercell_struct}
The crystal structures of the \fecofivep{} supercells. 
(a-c) Three configurations of Fe$_{4}$Co$_{1}$PB$_2$ and (d) single configuration for Fe$_{3}$Co$_{2}$PB$_2$.
}
\end{figure}
A supercell method was used to model the chemical disorder.~\cite{dane_density_2015,edstrom_magnetocrystalline_2017}
To build a supercell, multiplication of the basal unit cell and replacement of an appropriate amount of atoms of one type by atoms of the other type were made.
The Fe atoms were replaced by Co or 5$d$ atoms forming the \fecofivep{} and (Fe$_{0.95}$X$_{0.05}$)$_5$PB$_2$ compositions (X = 5$d$ element).
For \fecofivep{} the considered intermediate compositions were: $x$~=~0.2, 0.4, 0.6, and 0.8.
The MAE calculations based on the supercell method~\cite{dane_density_2015} are uncommon, as they are time-consuming even for relatively simple alloys.
The reason for that is significant increase in the number of inequivalent atomic positions generated for the supercell model.
Additionally, accurate results require averaging over several different large supercells.~\cite{dane_density_2015,steiner_calculation_2016}
It limits the size of supercells which we can use for MAE calculations.
Hence, we study only the supercells including symmetry operations and consisting of up to 16 inequivalent atoms.
The considered crystal structures are presented in Fig.~\ref{fig:supercell_struct}.
Three configurations were considered for Fe$_{4}$Co$_{1}$PB$_2$ and  Fe$_{1}$Co$_{4}$PB$_2$ and one for Fe$_{3}$Co$_{2}$PB$_2$ and Fe$_{2}$Co$_{3}$PB$_2$ compositions.
For the considered supercell models, the energy convergence with a number of $\mathbf{k}$-points was carefully tested.
%
%
The supercell method was also employed to calculate the MAE 
of (Fe$_{0.95}$X$_{0.05}$)$_5$PB$_2$ compositions
with various 5$d$ elements X.
To construct the models, 
one of twenty Fe atoms  in the basal \fefivep{} supercell was replaced by the dopant. 
It led to the crystal structures containing 10 inequivalent atomic positions.
For calculations of the systems with 5$d$ dopants a relatively dense $20 \times 20 \times 10$ \textbf{k}-mesh was used, in order to get the well converged results of MAE.

%
To compute the Curie temperatures within the mean-field theory ($T_\mathrm{C}^{\mathrm{MFT}}$) for the whole series of \fecofivep{} compositions the FPLO5.00 version of the code was used.~\cite{koepernik_self-consistent_1997}
The $T_\mathrm{C}^{\mathrm{MFT}}$ is proportional to the total energy difference between the ferromagnetic and paramagnetic configurations~\cite{gyorffy_first-principles_1985, sato_curie_2003, kudrnovsky_exchange_2004} 
according to:
\begin{equation}\label{eq:1}
k_\mathrm{B} T_\mathrm{C}^{\mathrm{MFT}} = \frac{2}{3} \frac{E_{\mathrm{DLM}} - E_{\mathrm{FM}}}{c},
\end{equation}
where $E_{\mathrm{DLM}}$ and $E_{\mathrm{FM}}$ are total energies for the paramagnetic and ferromagnetic configurations, 
$k_{\mathrm{B}}$ is Boltzmann constant, and
$c$ is total concentration of magnetic atoms. 
In case of \fefivep{} containing five Fe atoms (considered as magnetic ones) within a formula unit consisting of eight atoms, the concentration parameter $c$ is equal 5/8.
To model the paramagnetic state the disordered local moment (DLM) method was used,~\cite{heine_theory_1981}
in which the thermal disorder among the magnetic moments is modeled by using the coherent potential approximation (CPA).~\cite{soven_coherent-potential_1967}
The FPLO5 is the latest public version of the code allowing for the CPA calculations and does not have implemented the GGA.
Thus, the local density approximation (PW92)~\cite{perdew_accurate_1992} form of the exchange-correlation potential had to be chosen.
For the calculations within FPLO5, a scalar-relativistic mode and a $12 \times 12 \times 12$ \textbf{k}-mesh were used.
In the FPLO5 the magnetically ordered state (resulting in $E_{\mathrm{FM}})$ was artificially modeled within the CPA, to avoid numerical discrepancies between the ordered (in principle non-CPA) and DLM (CPA) models.
In calculations using the FPLO5 code, the minimum basis have been optimized for the terminal compositions \fefivep{} and \cofivep{}, 
subsequently the resultant compression parameters were used for intermediate compositions modeled with CPA.
%
%
The VESTA code was used for visualization of crystal structure.~\cite{momma_vesta_2008}

\subsection{Experimental Details}\label{subsec:exp_details}
%
%
The samples in the series \fecofivep{} ($x$ from 0.0 to 0.7) were synthesized by mixing stoichiometric amounts of the master alloys \fefivep{} and \cofivep{}.
The master alloys were prepared, in accordance with previous studies~\cite{hedlund_magnetic_2017}, 
from pure elements of 
iron (Leico Industries, purity 99.995\%, surface oxides reduced in H$_2$-gas), 
cobalt (Johnson Matthey, purity 99.999\%), 
phosphorus (Cerac, purity 99.999\%), 
and 
boron (Wacher-Chemie, purity 99.995\%). 
This was done by forming first the TM$_2$B (TM~=~Fe, Co), using a conventional arc furnace, and subsequently dropping the phosphorus in a melt of the metal boride in an induction furnace using the drop synthesis method.~\cite{carlsson_determination_1973} 
All samples were subsequently crushed, pressed into pellets, and heat treated in evacuated silica ampules at 1273~K for 14 days after which they were quenched in cold water.
At $x$ higher than 0.7 the correct crystalline phase could not be produced, 
all attempts resulted in a decomposition to other crystalline phases.

%
To study the phase content and to perform crystal structure analysis of all samples, a powder X-ray diffraction (XRD) was used. 
The measurements were done using a Bruker D8 diffractometer equipped with a LynxEye position sensitive detector (4\textdegree\ opening) using CuK$\alpha_1$ radiation ($\lambda$~=~1.540598~\AA{}) at 298~K in a 2$\theta$ range of 20\textdegree--90\textdegree. 
The crystal structures were evaluated with the software FullProf~\cite{rodriguez-carvajal_recent_1993} using refinements according to the Rietveld method.~\cite{rietveld_profile_1969} 
The unit cell parameters were precisely studied using the least square refinements of the peak positions, employing the software UnitCell.~\cite{holland_unit_1997}

%
The synthesized samples were magnetically studied using a Quantum Design PPMS 6000.
Samples were immobilized in gelatin capsules with varnish.
The magnetization at 3~K was measured between applied magnetic fields of 0 and 7.2~MA\,m$^{-1}$.
The magnetization in SI units was calculated from magnetic moment using the sample weight and the crystallographic volume obtained from the XRD measurements at 298~K. 
When approaching magnetic saturation the magnetization process is described by the law of approach to saturation (LAS).~\cite{chikazumi_physics_1997}
LAS has been formulated in several ways~\cite{chikazumi_physics_1997,andreev_law_1997,zhang_law_2010,brown_theory_1940},
but it takes a general form 
\begin{equation}
\frac{M}{M_{\mathrm{S}}} = \sum_{j} a_j H^j,
\end{equation}
where $j$ is usually an integer, $a_j$ are coefficients, $M$ and $M_{\mathrm{S}}$ are magnetization and saturation magnetization, and $H$ is the applied magnetic field.
The LAS was used to determine an effective anisotropy constant $\left| {K}_{\mathrm{eff}} \right|$ in the same implementation as we used before.~\cite{cedervall_magnetostructural_2016,hedlund_magnetic_2017}
The interval 93\%--98\% of the magnetic saturation was used.
The applied formula was 
\begin{equation}
\label{eq:MS}
\frac{M}{M_{\mathrm{S}}} = 1 + aH + \frac{b}{H} + \frac{c}{H^2}.
\end{equation}
The experimental data was fit with four models in which $a$ and $b$ coefficients can be zero or non-zero
and since $\frac{1}{H^2}$ term is used to extract $\left| {K}_{\mathrm{eff}} \right|$ this part is always considered as non-zero.
$\left| {K}_{\mathrm{eff}} \right|$ is given here by 
\begin{equation}
\label{eq:Keff}
\left| {K}_{\mathrm{eff}} \right| = \sqrt{ \frac{15c}{4} } \mu_0 M_{\textrm{S}}.
\end{equation}
The difference in results between all four models are relatively small (max. 0.20~MJ\,m$^{-3}$), 
thus in the experimental section we present only the $\left| {K}_{\mathrm{eff}} \right|$ for the simplest model with the coefficients $a$~=~$b$~=~0.
%

\section{Results and Discussion}\label{sec:results}

The results of first-principles calculations of technologically important magnetic parameters for the considered systems are shown.
For \fecofivep{} the $M_\mathrm{S}$, $T_\mathrm{C}$, and MAE are presented.
For (Fe$_{0.95}$X$_{0.05}$)$_5$PB$_2$ (X = 5$d$ element) the results are limited to MAE and partial magnetic moments.
For the main phase -- \fefivep{} -- a detailed analysis of  electronic structure, magnetic moments, Fermi surface, and MAE is given.
The theoretical efforts  are complemented by experimental synthesis and measurements of the considered \fecofivep{} compositions.
\subsection{Crystal Structure and Electronic Structure of Fe$_5$PB$_2$ and Co$_5$PB$_2$}\label{subsec:el_struct}

\begin{table}[!ht]
\caption{\label{tab:crystal_data} 
The optimized crystallographic parameters for Fe$_5$PB$_2$ and Co$_5$PB$_2$ as calculated with the FPLO14 code, using the GGA(PBE) functional, with (SP) and without (NM) spin polarization. 
Space group $I$4/$mcm$, no. 140. 
The Wyckoff positions are: Fe$_1$/Co$_1$ ($x$, $x$+1/2, $z$), Fe$_2$/Co$_2$ (0, 0, 0), P (0, 0, 1/4), and B ($x$, $x$+1/2, 0).
For comparison the values measured in this work at room temperature for Fe$_5$PB$_2$ and the literature values for Co$_5$PB$_2$ are also reported. 
}
\centering
\footnotesize
\begin{tabular}{l|cccccc}
\hline \hline
system        						& $a$ [\AA{}] & $c$ [\AA{}] & $x_{\text{Fe$_1$/Co$_1$}}$ & $z_{\text{Fe$_1$/Co$_1$}}$ & $x_{\text{B}}$ &  $c/a$\\
\hline
Fe$_5$PB$_2$ (GGA-SP) 							&   5.456     &   10.296    &   0.170          &   0.139          & 0.381  & 1.887\\ 
Fe$_5$PB$_2$ (expt.) 							&   5.492     &   10.365    &   0.170          &   0.141          & 0.381 & 1.887\\
\hline
Co$_5$PB$_2$ (GGA-SP)							&   5.284     &   10.541    &   0.169          &   0.142          & 0.376  & 1.995\\ 
Co$_5$PB$_2$ (GGA-NM)							&   5.309     &   10.406    &   0.169          &   0.141          & 0.376 & 1.960  \\ 
Co$_5$PB$_2$ (expt.)~\cite{rundqvist_x-ray_1962} 			&   5.42      &   10.20     &   -              &   -              & - & 1.882\\ 
\hline \hline
\end{tabular}
\end{table}

%
The optimized crystallographic parameters of \fefivep{} and \cofivep{} are compared in Table~\ref{tab:crystal_data} with the results of measurements.
For \fefivep{} the agreement between the GGA and experiment is good and for the \cofivep{} the GGA underestimates $a$ and overestimates $c$. 
The disagreement may originate from both theory and experiment.
The lattice parameters of \cofivep{} were last refined by Rundqvist back in 1962.~\cite{rundqvist_x-ray_1962} 
Unfortunately, we did not manage to synthesize the \cofivep{} sample.
According to comprehensive study of Haas~\textit{et al.} the PBE remains the best GGA functional for most of the solids containing 3$d$ transition elements.~\cite{haas_calculation_2009} 
However, it has a tendency to overestimate the lattice constants.~\cite{haas_calculation_2009}
The presented PBE results for \fefivep{} go against this trend.
PBE underestimates also a volume of \cofivep{}.
The observed underestimation of lattice parameters/volumes is similar to the results obtained from GGA for bcc Fe~\cite{ropo_assessing_2008} and fcc Co~\cite{zeleny_ab_2008}, 
for which the use of GGA leads to about 0.5 - 1.0\% underestimation of the lattice parameters (which is equivalent to about 1.5 - 3.0\% underestimation in volume). 
In case of \fefivep{} and \cofivep{} the calculated (with spin polarization) PBE volumes are 2.2 and 1.8\% underestimated, respectively.
It is surprising, however, that when the $c/a$ ratio for \fefivep{} is in agreement with experiment (both values are equal to 1.887), the corresponding result for \cofivep{} from (spin polarized) GGA is significantly different (1.995 against the experimental value 1.882).
The non-magnetic GGA calculations leads for \cofivep{} to $c/a$ equal to 1.960 -- also significantly different from the measured value.
We can only give a very general explanation for this discrepancy as coming from the insufficient treatment of corrections in PBE functional.

\begin{figure}[!ht]
\includegraphics[angle=270,trim = 75 0 35 0 ,clip,width=\columnwidth]{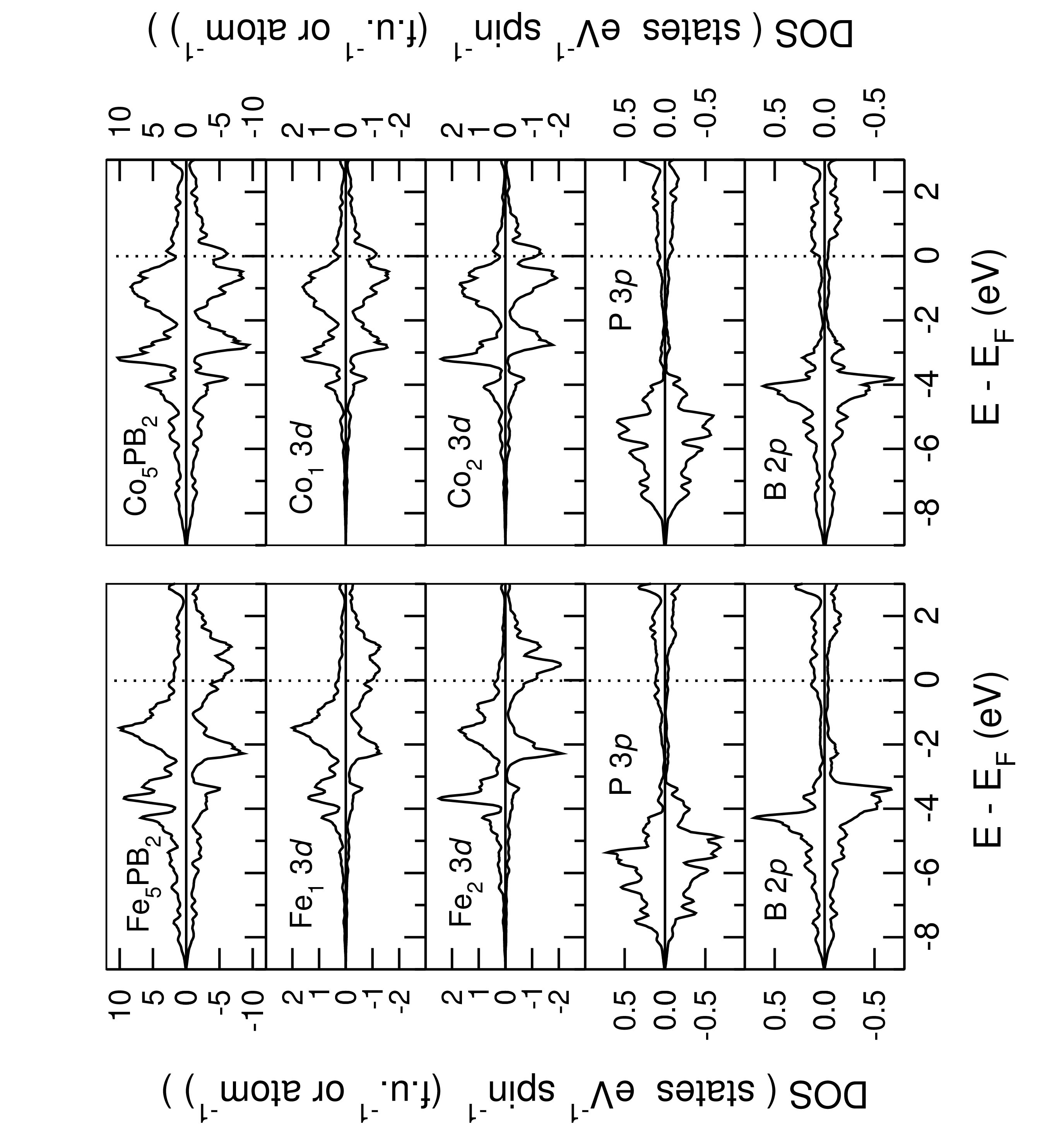}
\caption{\label{fig:dos} 
The spin projected partial and total densities of states (DOS) for Fe$_5$PB$_2$ and Co$_5$PB$_2$. 
Calculations were done within the FPLO14 code using the PBE functional and treating the relativistic effects in a full 4-component formalism (including spin-orbit coupling).
}
\end{figure}
The spin projected partial and total densities of states (DOS) for Fe$_5$PB$_2$ and Co$_5$PB$_2$ are presented in Fig.~\ref{fig:dos}.
The valence bands of these two metallic systems start around -9 eV.
In a range from -9 to -3~eV the main contributions to a valence band come from the P~3$p$ and B~2$p$ orbitals, while from -5~eV up to above $E_{\mathrm{F}}$ the dominant role play the 3$d$ orbitals.
The observed spin splitting (proportional to the magnetic moment) is bigger for Fe$_5$PB$_2$ than for Co$_5$PB$_2$,
which is related to a higher filling of the valence band for Co$_5$PB$_2$ than for Fe$_5$PB$_2$.
The majority spin channels of the two compounds are similar and nearly completely occupied.
The additional electrons in the Co$_5$PB$_2$ fill mainly the minority spin channel, reducing the magnetic moment.
The weak spin polarization of the P~3$p$ and B~2$p$ orbitals is induced by the 3$d$ orbitals.
The spin polarization on the Fermi level is defined as $P=|\frac{D_\mathrm{u}-D_\mathrm{d}}{D_\mathrm{u}+D_\mathrm{d}}|$, where $D_\mathrm{u}$ is the density of states at the Fermi level of the majority spin channel, and $D_\mathrm{d}$ for the minority spin channel.
The calculated spin polarization on the Fermi level (a total value including Fe, Co, P, and B contributions) is about 0.46 for Fe$_5$PB$_2$ and 0.60 for Co$_5$PB$_2$.

\subsection{Magnetic Moments of \fecofivep{}}\label{subsec:mag_mom}

%
\begin{figure}[ht]
\includegraphics[trim = 180 55 20 100,clip,height=\columnwidth,angle=270]{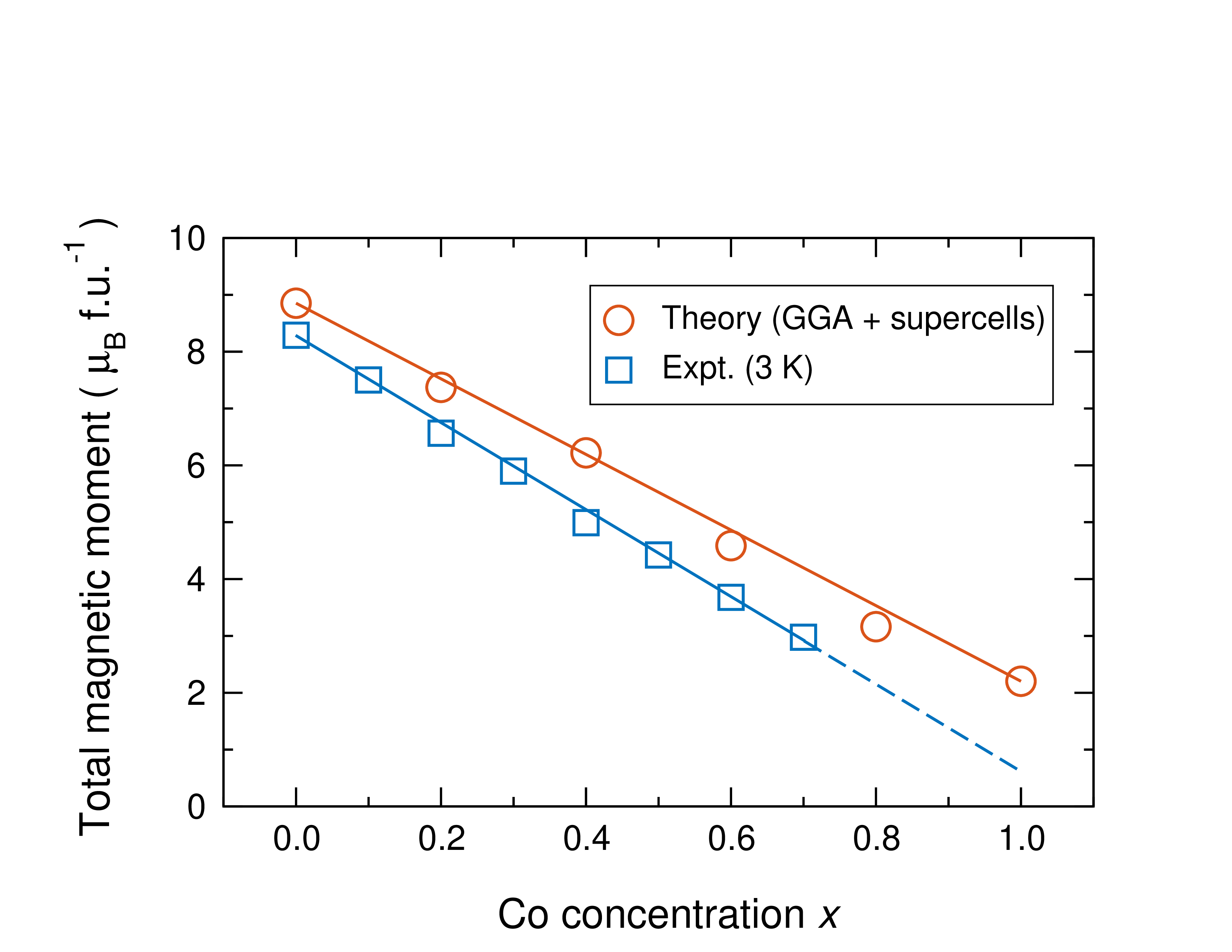}
\caption{\label{fig:m_vs_x} 
The Co concentration dependence of total magnetic moment for the \fecofivep{} system. 
The results calculated with supercell method are denoted by red circles,
the results measured at 3~K by blue squares.~\cite{cedervall_influence_2018}
Linear fits are drawn for a better perception.
Calculations were done with the FPLO14 code, using the GGA functional (PBE), and treating the relativistic effects in a full 4-component formalism (including spin-orbit coupling).
}
\end{figure}

\begin{table}[!htb]
\caption{\label{tab:mm} 
The spin, orbital, and total magnetic moments ($\mu_{\mathrm{B}}$ (atom or f.u.)$^{-1}$)  for Fe$_5$PB$_2$ and Co$_5$PB$_2$ as calculated along the quantization axis [001] (easy axis) with the FPLO14 code using the PBE functional and treating the relativistic effects in a full 4-component formalism (including spin-orbit coupling).
The saturation magnetization $M_\mathrm{S}$ (MA\,$m^{-1}$) is evaluated based on the total magnetic moments $m$ and theoretical lattice parameters.
}
\begin{tabular}{c|cc|cc}
\hline \hline
		&\multicolumn{2}{c|}{Fe$_5$PB$_2$}&\multicolumn{2}{c}{Co$_5$PB$_2$}\\
\hline       
site   		& $m_\mathrm{s}$   & $m_\mathrm{l}$     & $m_\mathrm{s}$   & $m_\mathrm{l}$\\
\hline
3$d_1$ 		&  1.78   &   0.033   &  0.41   &  0.011   \\
3$d_2$ 		&  2.11   &   0.052   &  0.64   &  0.013   \\
P	   		& -0.13   &   0.002   & -0.02   &  0.001   \\
B      		& -0.21   &   0.001   & -0.05   &  0.000   \\
\hline
$m$		&\multicolumn{2}{c|}{8.85}&\multicolumn{2}{c}{2.20}\\
$M_\mathrm{S}$ &\multicolumn{2}{c|}{1.07}&\multicolumn{2}{c}{0.28}\\
\hline \hline
\end{tabular}
\end{table}

The calculated Co concentration dependence of the total magnetic moment (a sum of spin and orbital contributions) for the \fecofivep{} system is presented in Fig.~\ref{fig:m_vs_x}
together with the experimental results at low temperature (3~K).~\cite{cedervall_influence_2018}
Whereas the results presented here are based on the supercell approach~\cite{cedervall_influence_2018},
in our previous work one can find the corresponding $m$($x$) plots based on the virtual crystal approximation (VCA) and coherent potential approximation (CPA).
The calculated and experimental $m$($x$) curves presented in Fig.~\ref{fig:m_vs_x} stay in good qualitative agreement, showing a linear decrease of magnetic moment with Co concentration.
Nevertheless, they differ by about 0.5 -- 1.0 $\mu_{\mathrm{B}}$/f.u., 
where the lower values come from measurements.
The reasons for this discrepancy should be sought on both experimental and theoretical sides.
Looking at the experiment, it is worth noting that the samples produced in this work are slightly non-stoichiometric and with a small amount of impurities.~\cite{cedervall_influence_2018}
Our measurements at 3~K for a powder sample of \fefivep{} showed a total magnetic moment equal to 8.29~$\mu_{\mathrm{B}}$/f.u.
in comparison to 8.6~$\mu_{\mathrm{B}}$/f.u. obtained by Lamichane \textit{et al.} for a \fefivep{} single crystal at 2~K.~\cite{lamichhane_study_2016}
It leads us to the conclusion that the magnetic moments we have measured may be slightly underestimated.
The calculated total magnetic moment of Fe$_5$PB$_2$ (8.85~$\mu_{\mathrm{B}}$/f.u.) using GGA is closer to the result obtained for the single crystal than for the powder sample. 
The discrepancy between the result of GGA calculations and single crystal measurements can then be attributed to the insufficiency of the GGA in description of correlations, however the calculations still provide an acceptable level of agreement with experiment.

%
%
The calculated spin, orbital, and total magnetic moments ($m_\mathrm{s}$, $m_\mathrm{l}$, $m$) for Fe$_5$PB$_2$ and Co$_5$PB$_2$ are collected in Table~\ref{tab:mm}.
For Fe$_5$PB$_2$ the calculated magnetic moments on Fe$_1$ and Fe$_2$ sites are equal to 1.81 (1.62) and 2.16 (2.16)~$\mu_{\mathrm{B}}$, respectively,
where in parentheses are given estimations from the magnetic hyperfine fields.~\cite{haggstrom_mossbauer_1975}
The induced spin magnetic moments on P and B are relatively small and oriented antiparallel to the dominant 3$d$ moments on Fe/Co.
The total magnetic moments of Fe$_5$PB$_2$ and Co$_5$PB$_2$ are almost entirely of spin character, where the 3$d$ orbital magnetic moments ($m_\mathrm{l}$'s) are nearly quenched.
The $m_\mathrm{l}$'s of Fe$_1$ and Fe$_2$ of the Fe$_5$PB$_2$ (calculated for the [001] quantization axis) are equal to 0.033~$\mu_{\mathrm{B}}$ and 0.052~$\mu_{\mathrm{B}}$, respectively.
These values surround the $m_\mathrm{l}$ value calculated for bcc Fe (0.043~$\mu_{\mathrm{B}}$) and are reduced in comparison to the experimental value for the bcc Fe (0.086~$\mu_{\mathrm{B}}$).~\cite{chen_experimental_1995}
The underestimation of the orbital magnetic moment in transition metals is recognized as a general weakness of the LDA and GGA.
Finally, almost no orbital contributions are observed for P and B atoms ($m_\mathrm{l} \sim 10^{-3}~\mu_{\mathrm{B}}$).
%
%
The calculated $m$ of Co$_5$PB$_2$ is equal to 2.20~$\mu_{\mathrm{B}}$/f.u. (0.44~$\mu_{\mathrm{B}}$/Co atom).
For comparison, the experimental magnetic moment of hcp Co is equal to 1.67~$\mu_B$/atom.~\cite{reck_orbital_1969}
The calculated $m_\mathrm{l}$'s of Co$_1$ and Co$_2$ of the Co$_5$PB$_2$ are equal to 0.011~$\mu_{\mathrm{B}}$ and 0.013~$\mu_{\mathrm{B}}$, respectively, and are one order of magnitude smaller than the $m_\mathrm{l}$ measured for hcp Co (0.13~$\mu_{\mathrm{B}}$).~\cite{moon_distribution_1964}
Although the theoretical values of magnetic moments for Co$_5$PB$_2$  have been presented above,
the magnetic ground state of this system has not been unambiguously resolved, 
which will be discussed in the next section.

\subsection{Curie Temperature of \fecofivep{}}\label{subsec:tc}

\begin{figure}[htb]
\centering
\includegraphics[trim=215 45 10 110,clip,height=\columnwidth,angle=270]{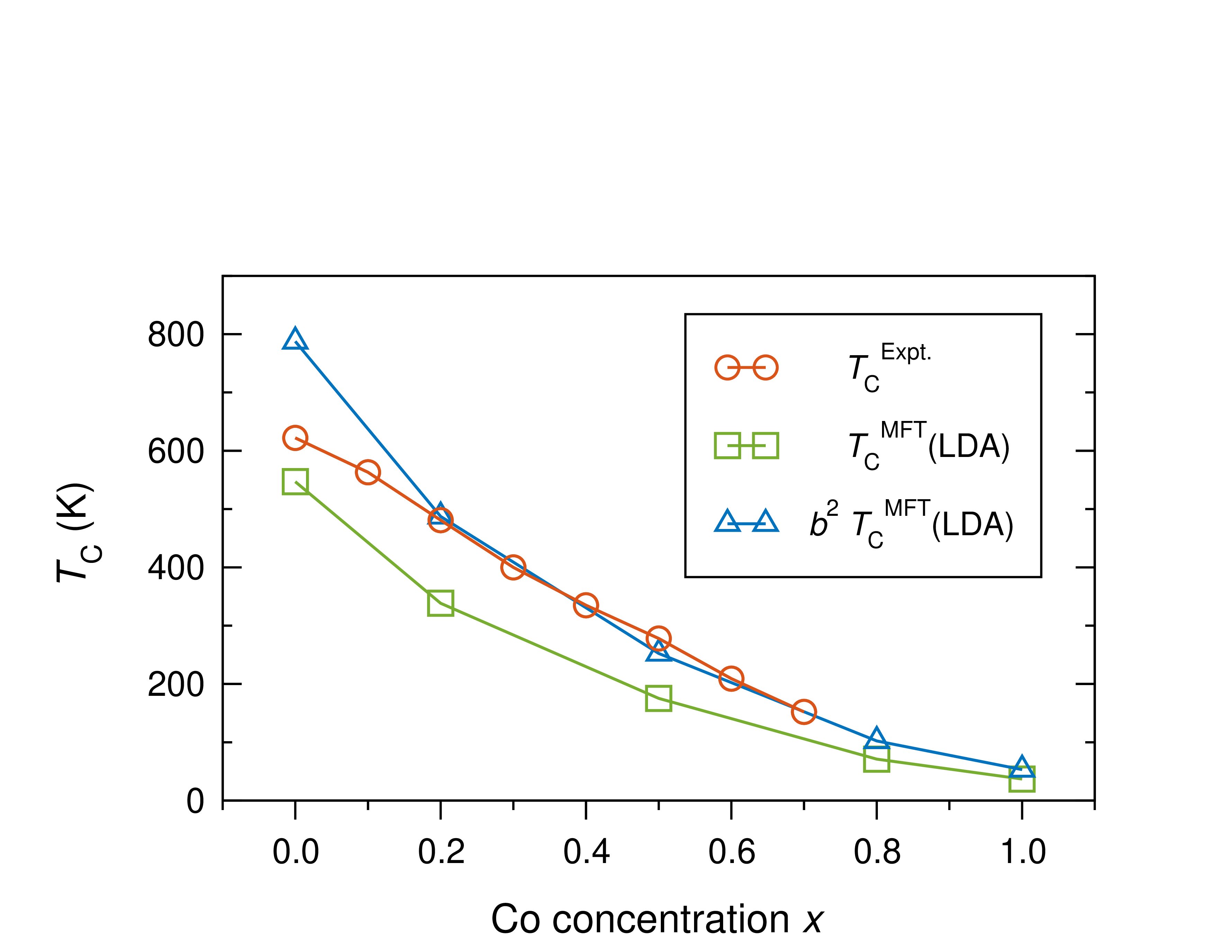} 
\caption{\label{fig:tc} 
The Curie temperatures as functions of Co concentration $x$ in \fecofivep{}.
The theoretical $T_\mathrm{C}^{\mathrm{MFT}}$(LDA) are calculated in the mean-field approximation with the FPLO5 code, 
using LDA functional, 
treating the chemical disorder with CPA, 
and modeling the paramagnetic state with DLM.
The experimental $T_\mathrm{C}$ was defined from the inflection point of field cooled magnetization \textit{versus} temperature measurements in a field of $\mu_0$H~=~0.01~T.~\cite{cedervall_influence_2018}
}
\end{figure}

The Curie temperatures ($T_\mathrm{C}^{\mathrm{MFT}}$(LDA)) calculated for the whole concentration range of the \fecofivep{} system within the mean-field theory and with the LDA functional are presented in Fig.~\ref{fig:tc}.
%
%
The observed overall decrease of the calculated $T_\mathrm{C}^{\mathrm{MFT}}$ with increase of Co concentration is consistent with experimental observations.~\cite{mcguire_magnetic_2015, cedervall_influence_2018} 
However, in the whole range in which it is possible to compare the MFT-LDA results with the experiment ($0.0 \leq x \leq 0.7$), theoretical values are smaller.
For example, the calculated $T_\mathrm{C}^{\mathrm{MFT}}$(LDA) of \fefivep{} is equal 547~K, whereas the corresponding experimental value is 622~K for the powder sample~\cite{cedervall_influence_2018}, or 655\textpm2~K for the single crystal.~\cite{lamichhane_study_2016}
%
%
This difference is due to the limitations of the MFT approach and insufficiency of the LDA in description of correlations.
By calculating Heisenberg exchange interactions, one could extract accurate critical temperatures using the random phase approximation (RPA) or Monte Carlo simulations.~\cite{sasioglu_above-room-temperature_2005,rusz_random-phase_2005}
The insufficiency of the LDA  manifests in underestimated values of the calculated magnetic moments of \fefivep{}; 7.30~$\mu_B$/f.u. \textit{versus} 8.6~$\mu_{\mathrm{B}}$/f.u. from experiment for a single crystal.~\cite{lamichhane_study_2016}
As it has been shown in the previous subsection, a much better description of magnetic moments of \fefivep{} in relation to the experimental result can be obtained by using the GGA functional instead LDA.
Thus, we suggest that the negative effect on $T_\mathrm{C}^{\mathrm{MFT}}$ coming from the limitations of the LDA can be partially corrected by using the correction parameter based on the magnetic moments obtained from GGA.
In Heisenberg model,  $T_\mathrm{C}^{\mathrm{MFT}}$ is proportional to squared effective moment ($m_\mathrm{eff}^2$).
Defining $b = \frac{m^\mathrm{GGA}}{m^\mathrm{LDA}}$ the \textit{corrected} Curie temperature is $b^2 T_\mathrm{C}^{\mathrm{MFT}}$(LDA), 
where in case of \fecofivep{} $b$ is about 1.2.
Figure~\ref{fig:tc} shows that for the region of intermediate Co concentrations the $b^2 T_\mathrm{C}^{\mathrm{MFT}}$(LDA) curve is in a better agreement with experiment than the uncorrected MFT-LDA results.

%
Unfortunately, we were unable to get experimental results of $T_\mathrm{C}$ for Co concentrations $x > 0.7$.
Because of that, we can not unambiguously resolve the issue of the magnetic ground state of terminal composition \cofivep{}.
Linear extrapolation of experimental magnetic moments for \fecofivep{} system suggests non-zero moment for \cofivep{}, see Fig.~\ref{fig:m_vs_x}.
On the contrary, linear extrapolation of the measured Curie temperature suggests a transition from ordered to disordered magnetic state at about $x = 0.9$, 
and therefore a non-magnetic ground state of \cofivep{}, see Fig.~\ref{fig:tc}.
Furthermore, experimental results reported by McGuire and Parker suggested absence of magnetic ordering for \cofivep{}.~\cite{mcguire_magnetic_2015}
From theoretical point of view, both uncorrected and corrected approaches show the non-zero values of $T_\mathrm{C}$ for Co-rich region ($T_\mathrm{C}^{\mathrm{MFT}}$(LDA)~=~37~K for \cofivep{}).
Taking into account (1) the problems with synthesis of the \cofivep{} phase, (2) preliminary character of the measurements reported by McGuire and Parker, (3) issues mentioned in previous subsection regarding optimization of the structural model of \cofivep{}, and (4) limitations of LDA/GGA in description of correlations of Co-rich phases, 
we conclude that 
based on existing data
the magnetic ground state of \cofivep{}
can not be definitively determined.

\subsection{Fermi Surface of \fefivep}\label{subsec:fs}
\begin{figure}[ht]
\includegraphics[width=\columnwidth]{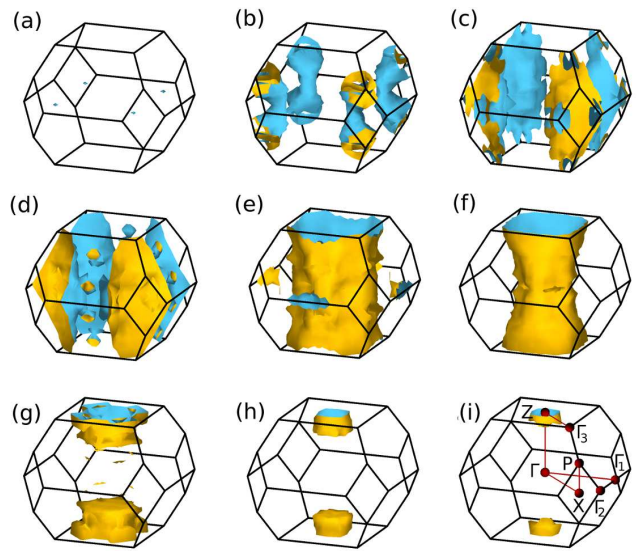}
\caption{\label{fig:fermi_surface} 
(a)--(i) The nine sheets of the Fermi surface of \fefivep{}. 
(i)
The $\mathbf{k}$-path used to calculate the band structure plot. 
Inside the visible tubes the sheets (e) and (f) contain the invisible pockets centered at $\Gamma$.
Calculations were done with the FPLO14 code using the PBE functional and treating the relativistic effects in a full 4-component formalism (including spin-orbit coupling).
}
\end{figure}
Figure~\ref{fig:fermi_surface} presents the calculated Fermi surface (FS) of \fefivep{} in a boundary of the first Brillouin zone.
The FS of \fefivep{} reflects the tetragonal symmetry of the crystal.
The FS consists of nine sheets and is relatively complex.
The states at the Fermi level ($E_{\mathrm{F}}$) have a Fe~3$d$ character, 
as can be read from the DOS plots in Fig.~\ref{fig:dos}.
The observed FS sheets can be divided into two groups.
The first group consists of four nested sheets of hole-type, see panels (a)--(d) of Fig.~\ref{fig:fermi_surface}, and
the second group includes the remaining five sheets of electron-type nested in a multiwalled way around the high symmetry point $Z$, see panels (e)-(i) of Fig.~\ref{fig:fermi_surface}.
While the sheets (c)--(f) form rather tubular shapes, allowing for open orbits along the symmetry axis, the remaining sheets, (a)--(b) and (g)--(i), take the form of pockets enabling only for closed FS orbits.~\cite{ziman_electrons_1962}
Because the band structure was calculated with spin-orbit coupling, 
the FS sheets cannot be unambiguously attributed to a particular spin channel.

\subsection{Magnetocrystalline Anisotropy of \fefivep}\label{subsec:MAE}

The results of investigating the MAE of \fefivep{} carried out in this work are: 
the band structure in vicinity of the Fermi level,
one- and two-dimensional $\mathbf{k}$-resolved MAE plots,
and the cross-section of FS.
Our inquiry is complemented by considerations of MAE engineering, as for example reduction of total magnetic moment.
%
%
The calculated MAE of the \fefivep{} is 0.52~MJ\,m$^{-3}$.
It indicates a uniaxial magnetocrystalline anisotropy with an easy axis along the tetragonal axis.
This result stays in a good agreement with the experimental value of anisotropy constant measured at 2~K (0.50~MJ\,m$^{-3}$) and with the previous theoretical findings (0.46~MJ\,m$^{-3}$).~\cite{lamichhane_study_2016}
Previously reported results for \fefivep{} show that $K_1$ first increases with temperature starting from 2~K up to about 100~K and then decreases to zero at $T_\mathrm{C}$.~\cite{lamichhane_study_2016}
The well known origin of the magnetocrystalline anisotropy is the spin-orbit coupling, which is taken into account in the fully relativistic full potential calculations.
In comparison with scalar relativistic approach, 
the fully relativistic one results in additional splitting of the electronic bands.
Since the spin-orbit coupling constant of $3d$-metals is of the order of 0.05~eV, the spin-orbit splitting also does not exceed this value.
%
%
The spin-orbit splitting leads to slightly different band structures for different quantization axes (e.g. for the orthogonal [001] and [100] axes).
%
\begin{figure*}[!ht]
\includegraphics[trim=260 25 15 30,clip,height=\textwidth,angle=270]{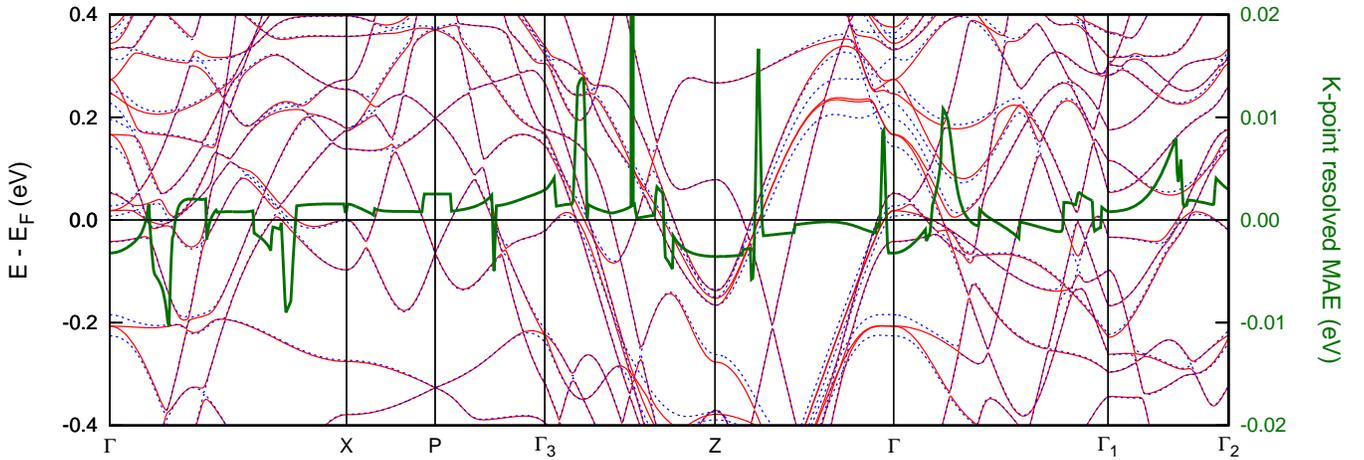}
\caption{\label{fig:fe5pb2_bands_MAE_narrow} 
The band structure of \fefivep{} calculated for quantization axes [100] (solid red lines) and [001] (dashed blue lines), 
together with the MAE contribution of each {\bf k}-point (thick green line) as obtained by the magnetic force theorem. 
The high symmetry points are presented within Brillouin zone in Fig.~\ref{fig:fermi_surface}~(i).
Calculations were done with the FPLO14 code using the PBE functional and treating the relativistic effects in a full 4-component formalism (including spin-orbit coupling).
}
\end{figure*}
Figure~\ref{fig:fe5pb2_bands_MAE_narrow} presents the band structures calculated for \fefivep{} in the proximity of $E_{\mathrm{F}}$,
together with the MAE contributions per $\mathbf{k}$-point obtained with the magnetic force theorem~\cite{liechtenstein_local_1987,wang_validity_1996,wu_spinorbit_1999} from the formula:
\begin{multline}\label{eq:10}
\mathrm{MAE} = E(\theta = 90^{\circ}) -  E(\theta = 0^{\circ}) =\\
= \sum_{\textrm{occ'}} \epsilon_{i}(\theta = 90^{\circ}) -  \sum_{\textrm{occ''}} \epsilon_{i}(\theta = 0^{\circ}), 
\end{multline}
%
where $\theta$ is an angle between the magnetization direction and the $c$ axis, $E(\theta)$ is a total energy for a specific direction; and $\epsilon_{i}$ is the band energy of the $i$th state.
The spin-orbit splitting is most easily observed for the energy window of a tenth eV around $E_{\mathrm{F}}$.
The $\mathbf{k}$-point resolved MAE takes positive and negative values, depending on the spin and orbital character of the bands near the Fermi energy. Generally, negative MAE-contributions coincide with occupied bands for a [100] spin quantization axis (solid red line) being pushed below corresponding bands for a [001] spin quantization axis (dashed blue line), and vice versa for positive contributions. For example, at the Z-point, there is a negative MAE contribution and at approximately -0.3~eV one can observe a solid red line below the dashed blue line. A more detailed analysis of the MAE contributions is in principle straight forward but somewhat complicated due to the complex band structure. Nevertheless, one can clearly observe the characteristic jumps where the bands cross $E_{\mathrm{F}}$, confirming the usual behavior that the MAE is determined by the electronic structure around the Fermi energy. Thus, controlling the MAE around $E_{\mathrm{F}}$ also allows for control of the MAE, as is practically 
possible, for example, via alloying. 
%
%

%
The same form of presentation of the $\mathbf{k}$-resolved MAE, as we have shown in Fig.~\ref{fig:fe5pb2_bands_MAE_narrow}, dominates in literature.
However, it is possible to plot the MAE($\mathbf{k}$) data within a three dimensional Brillouin zone, 
similar like the FS.
Recently, the 3D MAE($\mathbf{k}$) maps were presented for \fecotwo{} and FeNi.~\cite{belashchenko_origin_2015,werwinski_ab_2017}
\begin{figure}[ht]
\includegraphics[width=\columnwidth]{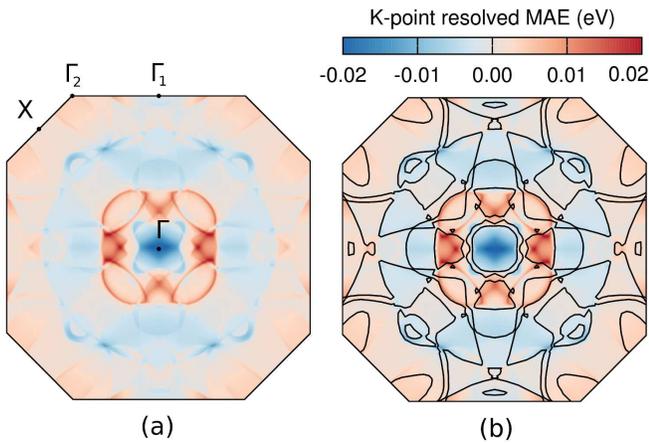}
\caption{\label{fig:fe5pb2_mae_vs_fs_slab} 
(a) The cross-section of the $\mathbf{k}$-resolved MAE with (b) the overlapped cross-section of the Fermi surface (black lines) for the \fefivep{}. 
The results of MAE($\mathbf{k}$) are obtained by the magnetic force theorem within 
the FPLO14 code using the PBE functional and treating the relativistic effects in a full 4-component formalism (including spin-orbit coupling).
}
\end{figure}
In Fig.~\ref{fig:fe5pb2_mae_vs_fs_slab}~(a) we show a cross-section of the MAE($\mathbf{k}$) (single plane going trough the $\Gamma$-point).
The selected profile is perpendicular to the easy axis [001], crosses the high symmetry point $\Gamma$, and is limited by the Brillouin zone boundaries.
The MAE($\mathbf{k}$) cross-section is a relatively complicated map of symmetric regions consisting of positive and negative contributions. 
The MAE contributions observed in Fig.~\ref{fig:fe5pb2_mae_vs_fs_slab} along the orthogonal axes [100] and [010] are not equal,
because the [100] direction is distinguished as quantization axis resulting in breaking of the four-fold symmetry. 
%
%
%
As the $E_{\mathrm{F}}$ is an upper integration boundary of total MAE, the FS sheets coincide with sharp changes in the $\mathbf{k}$-resolved MAE contributions.
It can be seen in Fig.~\ref{fig:fe5pb2_mae_vs_fs_slab}~(b), where the MAE($\mathbf{k}$) 2D plot is overlapped by the corresponding section of the FS.
As many of $\mathbf{k}$-resolved MAE contributions is in order of $10^{-3}$~eV per $\mathbf{k}$-point,
the total MAE value of about 10$^{-4}$~eV/f.u.  (83~$\mu$eV/f.u. or 0.52~MJ\,m$^{-3}$) indicates a fine compensation of many bigger components.
Unfortunately, this extra fine compensation and the complexity of the MAE($\mathbf{k}$) makes the ways to increase the MAE of the material difficult to predict.

\subsection{Fully Relativistic Fixed Spin Moment Calculations for \fefivep}\label{subsec:FSM}

%
The MAE value for \fefivep{} ($0.52$~MJ\,m$^{-3}$) is calculated with the equilibrium value of the magnetic moment (8.85~$\mu_{\mathrm{B}}$/f.u.). 
In the fixed spin moment (FSM) method~\cite{schwarz_itinerant_1984} the value of spin magnetic moment is considered as a parameter.
The fully relativistic implementation of FSM method allows to calculate the MAE as a function of spin magnetic moment.
Previously, we presented the MAE results as a function of FSM and Co concentration for the \fecotwo{} alloys.~\cite{edstrom_magnetic_2015} 
\begin{figure}[htb]
\centering
\includegraphics[trim = 165 40 35 100,clip,height=\columnwidth,angle=270]{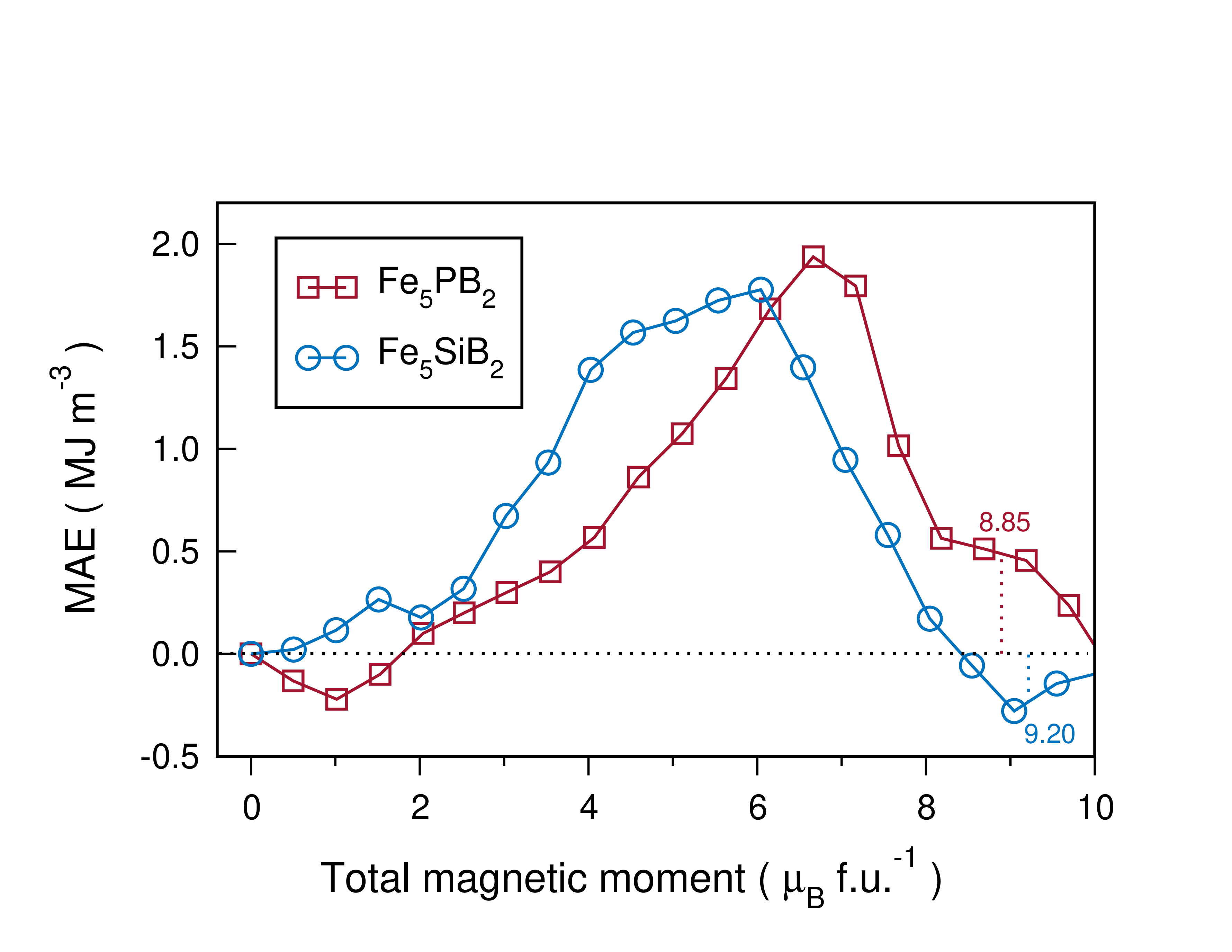}
\caption{
MAE as function of total magnetic moment ($m_{S}$+$m_{L}$) for \fefivep{} and \fefivesi{}~\cite{werwinski_magnetic_2016} as calculated with fixed spin moment (FSM) method with 
the FPLO14 code using the PBE functional and treating the relativistic effects in a full 4-component formalism (including spin-orbit coupling).
The equilibrium values of magnetic moments are denoted with dotted lines.}
\label{MAE_vs_fsm}
\end{figure}
Figure~\ref{MAE_vs_fsm} presents the evolution of the MAE with the total magnetic moment $m$ for the \fefivep{},
together with the previous results for \fefivesi{}.~\cite{werwinski_magnetic_2016}
The two MAE($m$) plots are similar in shape.
Going down from an equilibrium $m$ the corresponding MAE first increases, then it reaches maximum, to decrease finally to zero at $m$ equals zero. 
For \fefivep{} the maximum MAE($m$) is 1.94~MJ\,m$^{-3}$ 
for a fixed total magnetic moment of 6.7~$\mu_{\mathrm{B}}$/f.u.,
which means that the optimal magnetic moment has to be reduced by about 25\% with respect to the equilibrium value (8.85~$\mu_{\mathrm{B}}$/f.u.).
%
%
%
Thus, the question arises, how to stabilize this reduction.
A simple solution would be alloying the magnetic Fe by a non-magnetic element,
which often results in a linear decrease of magnetization.
However, alloying with a new element can severely affect the band structure,
which would change also the expected value of the MAE.
The smallest impact on the electronic structure should have 
substitutions chemically most similar to Fe and for this purpose we suggest Ru and Os of the Fe group.
Another strategy could be alloying of Fe ($Z_{\mathrm{Fe}} = 26$) with two elements at the same time, e.g. Cr ($Z_{\mathrm{Cr}} = 24$) and Ni ($Z_{\mathrm{Ni}} = 28$), keeping a constant number of the valence electrons, which should affect the band structure the least.
The above considerations, however, take into account only the band structure and neglect further issues like the crystal structure and size of the atoms, for example.

\subsection{Magnetocrystalline Anisotropy of \fecofivep}\label{subsec:MAE}
%
%
The effect of Fe/Co alloying on the MAE is not obvious in advance,
whereby the first-principles calculations are of great value in predicting the results, as has been shown previously for the \fecotwo{}~\cite{belashchenko_origin_2015} and \fecofivesi{}~\cite{werwinski_magnetic_2016} alloys.
%
\begin{figure}
	\centering
	\includegraphics[trim = 240 35 15 100,clip,height=\columnwidth,angle=270]{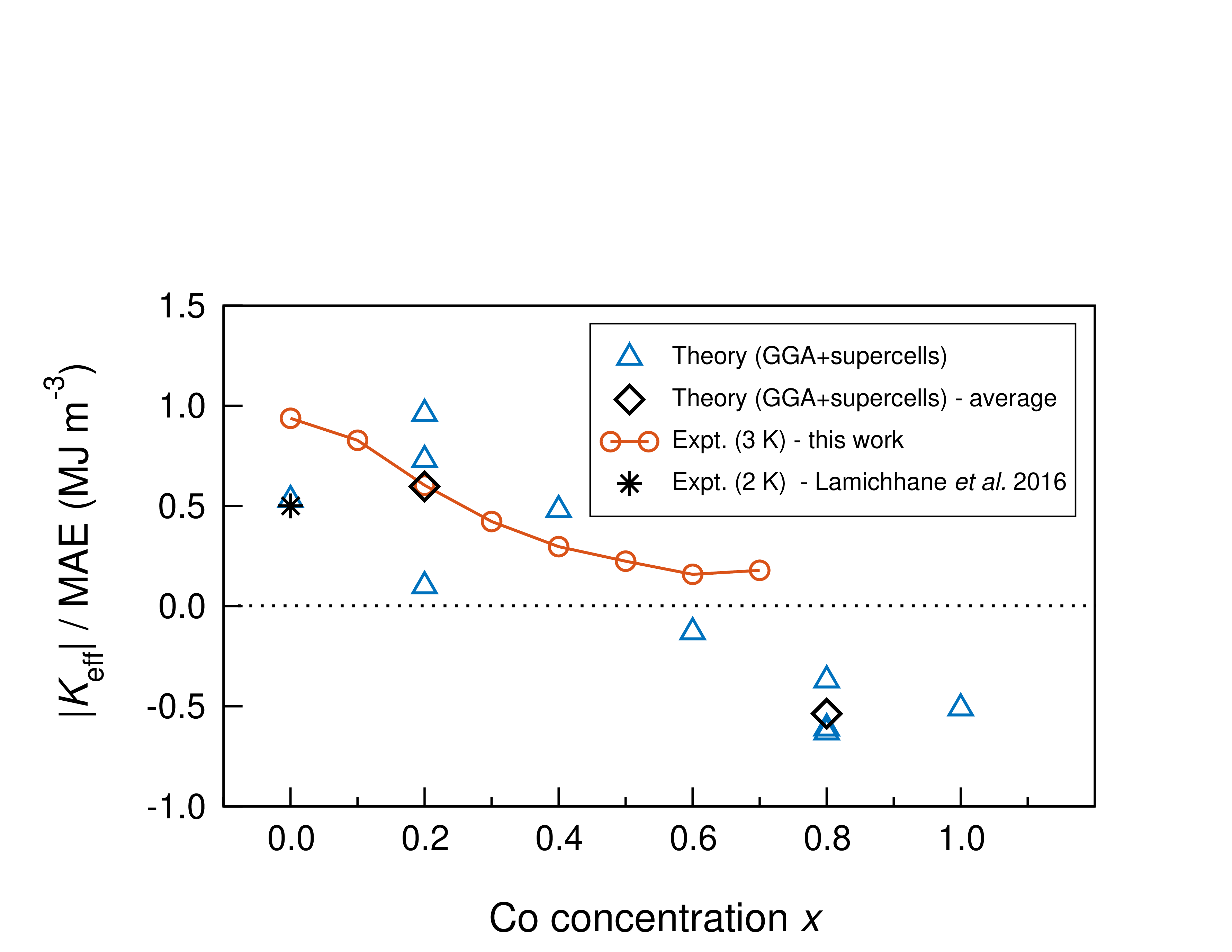}
	\caption{
	The experimental effective anisotropy constant $\left| K_{\textrm{eff}} \right|$ of the \fecofivep{} system at 3~K 
	(the sign of $K_{\textrm{eff}}$ is not considered), 
	 together with the magnetocrystalline anisotropy energy MAE values as calculated with the FPLO14 code.
	In calculations the supercell method for modeling of chemical disorder and the PBE functional were used.
	The relativistic effects were treated in a full 4-component formalism (including spin-orbit coupling).
	(for $x$ equal to 0.2 and 0.8 several inequivalent supercells are considered).
	For comparison the value of $K_1$ measured by
	 Lamichhane et al. for \fefivep{}.~\cite{lamichhane_study_2016}
	}
	\label{fig:fecopb2_Keff}
\end{figure}
Figure~\ref{fig:fecopb2_Keff} presents the MAE($x$) dependence for the \fecofivep{} system as calculated with use of the supercell method.
The MAE calculations based on the supercell method proved to be one of the most accurate method for evaluation the MAE.~\cite{dane_density_2015}
However, our calculations were limited by computational challenges of the supercell method.
Thus, in practice we were able to consider only a relatively small number of configurations, see Sec.~\ref{subsec:comp_details}.
The scattering of individual data points for $x = 0.2$ and $x = 0.8$ is in a similar range as observed by D\"ane~\textit{et al.}~\cite{dane_density_2015}
or Steiner~\textit{et al.}~\cite{steiner_calculation_2016} and shows that an averaging for several configurations is needed for accurate results. 
In Fig.~\ref{fig:fecopb2_Keff} the regions of positive and negative MAE (of perpendicular and in-plane anisotropy) are separated at Co concentration $x\simeq0.5$.
The calculated MAE is equal 0.52~MJ\,m$^{-3}$ for \fefivep{} and -0.51~MJ\,m$^{-3}$ for \cofivep{}.
Whereas, the anisotropy value close to zero, observed for $x\simeq0.5$, indicates a good soft magnetic material.
Figure~\ref{fig:fecopb2_Keff} presents also the low temperature measurements of the effective anisotropy constant $\left| {K}_{\mathrm{eff}} \right|$ carried out at 3~K for several \fecofivep{} compositions within the boundaries of $0.0 \leq x \leq 0.7$.
The value of $\left| K_{\textrm{eff}} \right|$ is the highest (0.94~MJ\,m$^{-3}$) for \fefivep{} and the the lowest for a Co concentration $x\sim0.6$.
$\left| K_{\textrm{eff}} \right|$ measured for \fefivep{} is significantly larger than the $K_1$~=~0.5~MJ\,m$^{-3}$ measured at 2~K for the single crystal.~\cite{lamichhane_study_2016}  
The decrease of $\left| K_{\textrm{eff}} \right|$ with $x$ is in agreement with the previous measurements for (Fe$_{0.8}$Co$_{0.2}$)$_5$PB$_2$ suggesting that 20\% Co substitution reduces the anisotropy field.~\cite{mcguire_magnetic_2015}
Previously we also showed the corresponding $\left| {K}_{\mathrm{eff}} \right|$ results for the Fe$_5$Si$_{1-x}$P$_{x}$B$_2$ system.~\cite{hedlund_magnetic_2017}
The presented values of $\left| K_{\textrm{eff}} \right|$ for \fefivep{} were $\sim$0.9~MJ\,m$^{-3}$ at 10~K and $\sim$0.65~MJ\,m$^{-3}$ at 300~K.~\cite{hedlund_magnetic_2017}
Notice that LAS is unable to determine the sign of $\left| K_{\textrm{eff}} \right|$ and thus the negative values of MAE predicted for $x \gtrsim 0.6$ cannot be confirmed by this method. 
Other methods, such as magnetometry measurements in different directions for single crystals or torque magnetometry would be preferable. 
Here, single crystals were not available, and up to 10 wt\% of impurities were present in the samples. 
Therefore, given the limitation in the model and the starting material the results presented from these should be seen as semi-quantitative. 
Taking into account the limitations of the LAS and the supercell method, the differences between theoretical and measured MAE($x$) results are acceptable.
We conclude, that Co alloying of \fefivep{} is not a good strategy to increase the MAE of this system.

\begin{figure}
	\centering
	\includegraphics[trim = 25 30 25 30,clip,width=1.0\columnwidth]{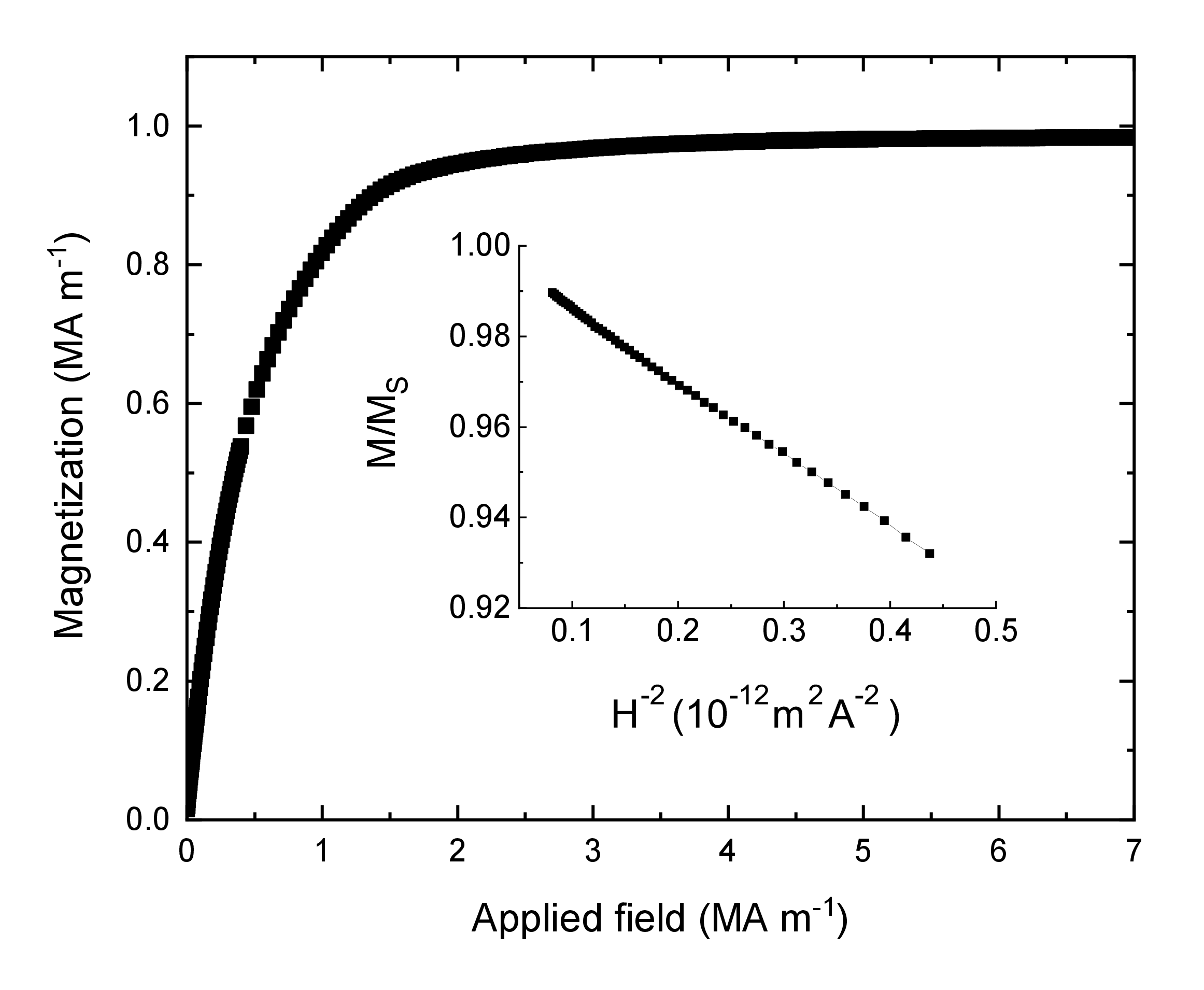}
	\caption{
Magnetization ($M$) as a function of applied field ($H$) measured for \fefivep{} at 3~K.
The inset shows a normalized magnetization ($M/M_\mathrm{S}$) as a function of 1/$H^2$.
	}
	\label{fig:fecopb2_m_vs_h}
\end{figure}
A typical magnetization ($M$) \textit{versus} applied field ($H$) curve measured at 3~K is shown in Fig.~\ref{fig:fecopb2_m_vs_h}.
The inset of Fig.~\ref{fig:fecopb2_m_vs_h} presents a plot of $M/M_\mathrm{S}$ \textit{versus} 1/$H^2$ as used to determine the $\left| {K}_{\mathrm{eff}} \right|$ within the LAS method.
More details on the implementation of the LAS method can be found in Sec.~\ref{subsec:exp_details}.

\subsection{Doping Fe$_5$PB$_2$ with 5$\boldsymbol{d}$ Elements}\label{subsec:5d}
One of the methods of tailoring the MAE is doping with 5$d$ elements.~\cite{edstrom_magnetic_2015,khan_site_2018}
Previously, we have confirmed that the 5$d$ elements can significantly affect the MAE due to a large spin-orbit coupling.~\cite{edstrom_magnetic_2015}
From the Fe$_5$Si$_{1-x}$P$_{x}$B$_2$ and \fecofivep{} systems, the highest MAE is found in the \fefivep{} phase.~\cite{hedlund_magnetic_2017}
Thus, it is considered as the parental compound for a further MAE engineering.
The MAE of (Fe$_{0.95}$X$_{0.05}$)$_5$PB$_2$ compounds (X~=~5$d$ elements) is calculated using the supercell method.
%
%
\begin{figure}
	\centering
	\includegraphics[trim = 30 75 250 110,clip,height=\columnwidth,angle=270]{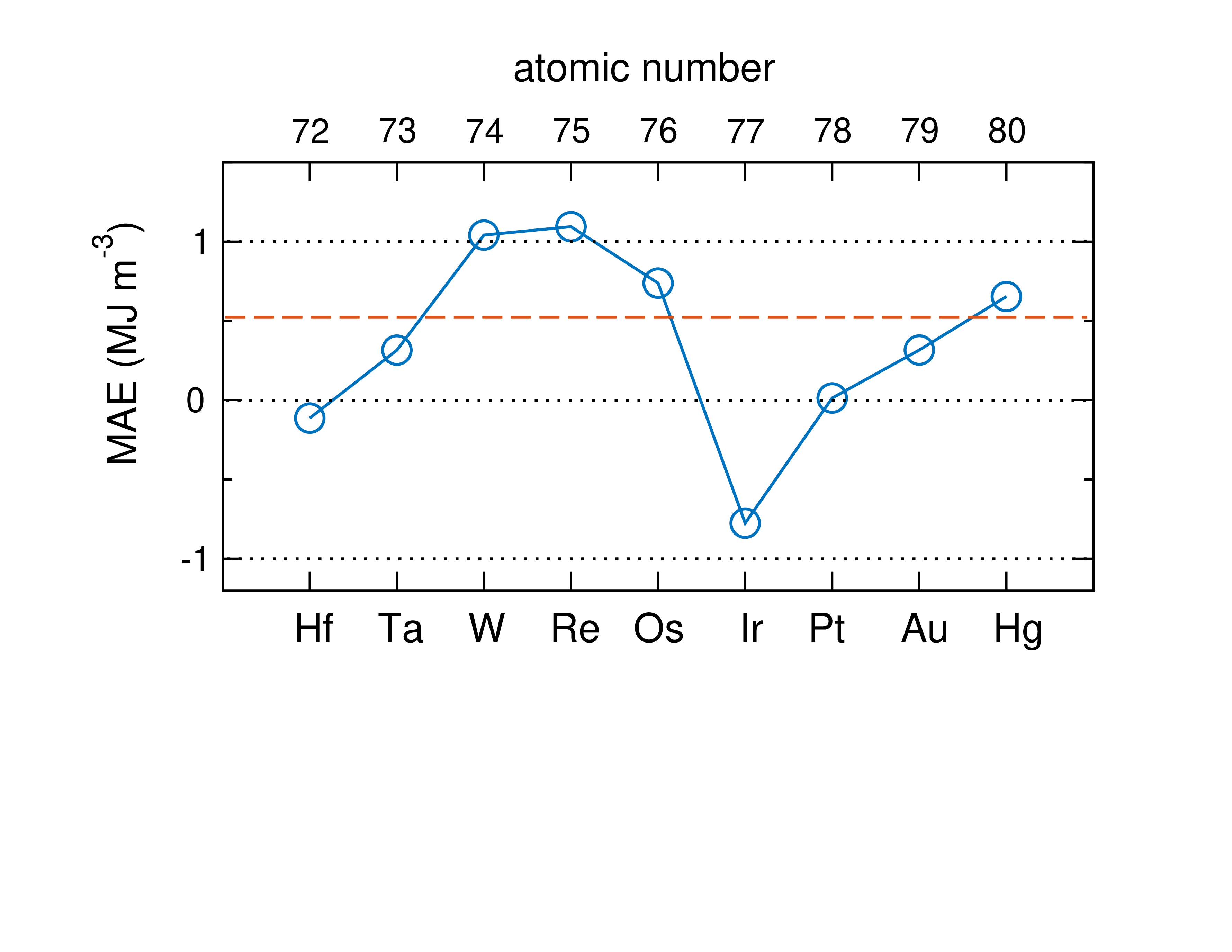}
	\caption{
MAE for various 5$d$ elements X in (Fe$_{0.95}$X$_{0.05}$)$_5$PB$_2$ as calculated with supercell method.
Calculations were done with the FPLO14 code using the PBE functional and treating the relativistic effects in a full 4-component formalism (including spin-orbit coupling).
The dashed line indicates the MAE of \fefivep{} (0.52~MJ\,m$^{-3}$) for comparison. 
	}
	\label{fig:fepb2_5d_MAE}
\end{figure}
The results are shown in Fig.~\ref{fig:fepb2_5d_MAE}, with the 5$d$ element marked on the $x$ axis and dashed line indicating the MAE of undoped Fe$_5$PB$_2$.
The 5$d$ doping has sometimes beneficial and sometimes adverse effect on MAE.~\cite{ayaz_khan_magnetocrystalline_2016, edstrom_magnetocrystalline_2017,edstrom_theoretical_2016}
Significant increase of MAE is observed for W or Re doping, similar like in the case of (Fe$_{1-x}$Co$_{x}$)$_2$B alloys investigated experimentally in our previous work~\cite{edstrom_magnetic_2015}.
The MAE grows from 0.52~MJ\,m$^{-3}$ for Fe$_5$PB$_2$ to about 1.1~MJ\,m$^{-3}$ for the compositions with W or Re, with 5\% Fe substitution.
Previously we have shown, that the increase in MAE observed for W and Re dopants is mainly due to the strong spin-orbit coupling of the 5$d$ atoms, however other variations in electronic structure also affect the MAE.~\cite{edstrom_magnetic_2015}
%
%
\begin{figure}
	\centering
	\includegraphics[trim = 30 55 255 110,clip,height=\columnwidth,angle=270]{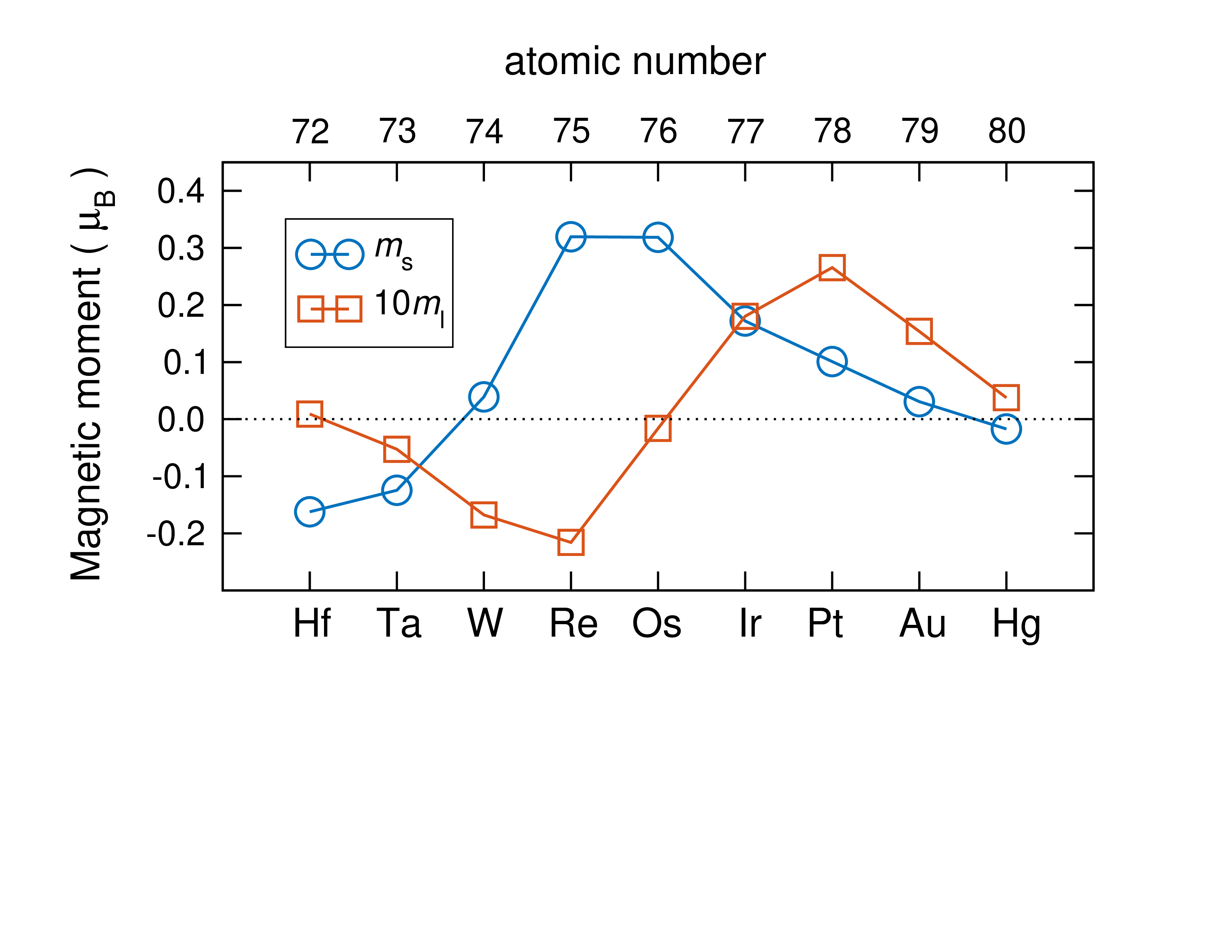}
	\caption{
	Spin ($m_\mathrm{s}$) and orbital  ($m_\mathrm{l}$) magnetic moments of 5$d$ transition metal impurities X in (Fe$_{0.95}$X$_{0.05}$)$_5$PB$_2$ as calculated for spin quantization axis along the $c$-axis.
	Supercell calculations were done with the FPLO14 code using the PBE functional and treating the relativistic effects in a full 4-component formalism (including spin-orbit coupling).
	}
	\label{fig:fepb2_5d_mm}
\end{figure}
Although in our calculations the 5$d$ elements are initially considered as non-magnetic, the dopants undergo spin polarization in a ferromagnetic medium and contribute to the total magnetic moment of the system.
The calculated spin and orbital magnetic moments on 5$d$ impurity show clear trend along the increasing atomic number of 5$d$ element, see Fig.~\ref{fig:fepb2_5d_mm}.
The spin magnetic moment of 5$d$ impurities are antiparallel to the Fe moments in the early 5$d$ series, while they are parallel in the late 5$d$ series.
Corresponding trends for 5$d$ atoms in magnetic 3$d$ hosts have been found previously computationally~\cite{akai_nuclear_1988, dederichs_ab-initio_1991} and experimentally.~\cite{wienke_determination_1991}

\section{Summary and Conclusions}\label{sec:conclulsions}
%
%
Our considerations began with a detailed theoretical analysis of the \fefivep{} compound.
The Fe 3$d$ orbitals are dominant in the valence band and responsible for the formation of large magnetic moments.
For the \fefivep{}
the fully relativistic band structure in the vicinity of Fermi level
was considered
to better understand the origin of the high value of magnetocrystalline anisotropy energy (MAE).
The calculated Fermi surface requires experimental confirmation.
%
%
The results of fully relativistic fixed spin moment calculations suggested that reduction of the magnetic moment of \fefivep{} should induce about fourfold increase of the MAE.
For practical realization of magnetic moment reduction it is suggested to alloy Fe with a non-magnetic element Ru or Os from the Fe group,
or to partially replace Fe with two elements at once, Cr and Ni, for example, keeping constant number of valence electrons.

%
Three critical parameters for technological applications: saturation magnetization ($M_\mathrm{S}$), Curie temperature ($T_\mathrm{C}$), and MAE
were calculated
for the whole concentration range between \fefivep{} and \cofivep{}.
The calculated $M_\mathrm{S}$ and $T_\mathrm{C}$ decreased with Co concentration and for the terminal composition \cofivep{} a weakly ordered magnetic ground state was predicted.
The calculated  $M$($x$) and $T_\mathrm{C}$($x$) were in decent agreement with the measurements, although the ferromagnetic ground state of \cofivep{} is questionable.
The Co doping in \fecofivep{} system gives the possibility of tuning the $T_\mathrm{C}$ in a range from about six hundred kelvins to almost down to zero.
The calculated MAE was positive for \fefivep{}, negative for \cofivep{}, and 
went through zero around 50\% Co concentration.
%
%
This picture of MAE($x$) behavior was in overall agreement with the experimental study of the effective anisotropy constant $\left| K_{\textrm{eff}} \right|$ for the \fecofivep{} alloys.
The measurements showed the highest $\left| K_{\textrm{eff}} \right|$ value for stoichiometric \fefivep{} which decreased with Co doping.
We concluded then that Co alloying is not a good strategy to increase the MAE of \fefivep{} alloy.
The measured $\left| K_{\textrm{eff}} \right|$ of about 0.94~MJ\,m$^{-3}$ at 3~K was, however, the highest value obtained so far for \fefivep{}, giving a hope for potential application of its other alloys.
It was also calculated how the 5\% doping of Fe with 5$d$ elements affects the MAE of the \fefivep{}.
It was shown that \fefivep{} doping with W or Re results in significant increase of the magnetocrystalline anisotropy energy.

\begin{acknowledgments}
MW and JR acknowledge the financial support from the Foundation of Polish Science grant HOMING. 
The HOMING programme is co-financed by the European Union under the European Regional Development Fund.
JC and MS acknowledge the financial support from the Swedish Research Council.
DH, PS, KG acknowledge Swedish Foundation for Strategic Research for financial support. 
Part of the computations were performed on resources provided by the Poznań Supercomputing and Networking Center (PSNC).
We thank Bartosz Wasilewski for help with language editing and Dr Jakub Kaczkowski for reading the manuscript and helpful discussion. 
\end{acknowledgments}

\end{sloppypar}

\bibliography{feco5pb2_MAE}        

\end{document}